# Out-of-equilibrium interactions and collective locomotion of colloidal spheres with squirming of nematoelastic multipoles

Bohdan Senyuk[a,b], Jin-Sheng Wu[a] and Ivan I. Smalyukh[a,b,c,d,*]

[a]Department of Physics, University of Colorado, Boulder, CO, USA.
[b]International Institute for Sustainability with Knotted Chiral Meta Matter (WPI-SKCM²), Hiroshima University, Higashi-Hiroshima, Hiroshima, Japan.
[c]Materials Science and Engineering Program, University of Colorado, Boulder, CO, USA.
[d]Renewable and Sustainable Energy Institute, National Renewable Energy Laboratory and University of Colorado, Boulder, CO, USA

*E-mail: ivan.smalyukh@colorado.edu

**Abstract**

Many living and artificial systems show a similar emergent behavior and collective motions on different scales, starting from swarms of bacteria to synthetic active particles, herds of mammals and crowds of people. What all these systems often have in common is that new collective properties like flocking emerge from interactions between individual self-propelled or externally driven units. Such systems are naturally out-of-equilibrium and propel at the expense of consumed energy. Mimicking nature by making self-propelled or externally driven particles and studying their individual and collective motility may allow for deeper understanding of physical underpinnings behind the collective motion of large groups of interacting objects or beings. Here, using a soft matter system of colloids immersed into a liquid crystal, we show that resulting so-called nematoelastic multipoles can be set into a bidirectional locomotion by external periodically oscillating electric fields. Out-of-equilibrium elastic interactions between such colloids lead to collective flock-like behaviors, which emerge from time-varying elasticity-mediated interactions between externally driven propelling particles. The repulsive elastic interactions in the equilibrium state can be turned into attractive interactions in the out-of-equilibrium state under applied electric fields. We probe this behavior at different number densities of colloidal particles and show that particles in a dense dispersion collectively select the same direction of a coherent motion due to elastic interactions between near neighbors. In our experimentally implemented design, their motion is highly ordered and without clustering or jamming often present in other colloidal transport systems, which is promising for technological and fundamental-science applications, like nano-cargo transport, out-of-equilibrium assembly and microrobotics.

**Introduction**

Emergent collective behavior of constituent units in different systems is a thriving interdisciplinary area of research. Researchers aspire to establish a detailed description of the collective motion of living organisms, self-propelled or driven objects on different scales. Collective motion involves numerous interacting units that perform coherent motion (1-3). The motility of the units can be self-generated as in living organisms (4), chemically-driven (5) or externally induced as in different colloidal systems (6-14). Developments in fabrication techniques allow for design of numerous synthetic colloids that exhibit a coherent propulsion in isotropic suspensions (5,8,15-17). As one of the fascinating types of active matter, some microorganisms are known to be roughly spherical in their shape, yet capable of swimming by continuously asymmetrically perturbing/deforming their shape, say via beating arrays of cilia on their surface; an example of such a microorganism is a ciliate like *Opalina* (18). A squirming model inspired by this behavior has been used to understand various aspects of the biological locomotion, where nonreciprocal surface profile evolution with respect to the static quasispherical shape gives origins to the self-propulsion.

In nematic colloidal systems, colloidal particles with various surface boundary conditions are known for inducing multipolar elastic deformations of the molecular alignment field, ranging from dipoles and quadrupoles to hexadecapoles and even higher-order multipoles (19-26), which pre-determine the nature of energy-minimizing elasticity-mediated colloidal interactions under equilibrium conditions. Reconfigurations of colloidal nematoelastic multipoles were previously demonstrated, allowing one to switch from attractive to repulsive forces between colloidal microparticles (27), back and forth, as well as cause colloidal spinning (28). Could controlled morphing of such particle-induced deformations of the aligned director field lead to colloidal particle propulsion in the nematic fluid hosts? If so, how would the out-of-equilibrium elasticity-mediated interactions between such spheres differ from their equilibrium counterparts, including their reconfigurations by external stimuli?

This article focuses on electric-field induced out-of-equilibrium dynamic properties and collective behavior of colloidal spheres immersed into a nematic liquid crystal (LC), which is a fluid with properties anisotropic with respect to a LC director $\mathbf{n}(\mathbf{r})$ describing the average local orientation of rodlike LC molecules (29). We study these colloid-LC dispersions experimentally in a homeotropic (boundary conditions for $\mathbf{n}(\mathbf{r})$ are perpendicular to the confining substrates) nematic LC cell and show that a bi-directional locomotion of colloidal particles can be activated by driving periodic nonreciprocal evolutions of the director field around them in response to the oscillating electric field applied between confining substrates with transparent electrodes. This results in a squirming motion of the colloids resembling the squirming motion of quasi-particle-like topological solitons in chiral LCs and other soft matter systems (30-34). In recent decades, LC's orientational elasticity mediated interactions between colloids were studied extensively under equilibrium conditions (12,19,20,23-26,35-44), whereas here we show that an oscillating electric field can cause an out-of-equilibrium evolution of elastic multipoles and a switch from repulsive to attractive elastic interactions between colloids, which would typically only repel under equilibrium conditions (19,23,25,41). Elastic interactions between particles emerge to reduce the free energy cost due to $\mathbf{n}(\mathbf{r})$ distortions around them, albeit here this happens without the dynamic $\mathbf{n}(\mathbf{r})$ fully reaching equilibrium because of the voltage modulation and ensuing colloidal motions. Such dynamic colloidal elastic multipoles with reconfigurable nature of induced elastic multipole moments can mutually repel or attract, where anisotropic interactions between them can be tuned by changing the voltage driving scheme. We probe the behavior of colloidal particles at the different number density and find that, at the higher number density, they show coherent uni-directional motion. The transition from bi-directional locomotion in dilute dispersions to a uni-directional locomotion in dense dispersions is caused by particles "sensing" the presence of their active-particle neighbors through interactions mediated by orientational LC elasticity. Under the applied electric field in dense dispersions, colloids collectively select the same direction of locomotion without clustering, jamming or clogging as in other colloidal systems (45,46). The direction of collective motion of colloids can be changed with respect to the predefined tilt of the director, which can be used for the development of different driven, active (47,48), and other out-of-equilibrium self-reconfigurable systems, microrobotics (49,50) and many other technological applications. Our findings also provide new insights into colloidal LCs and their assembly, out-of-equilibrium and collective behavior, as well as may enable new technological uses of nematic colloids.

**Results**

**Colloidal Spheres in a Homeotropic Nematic Cell.** Before proceeding with collective out-of-equilibrium effects in nematic colloids, we first study them as isolated objects under energy-minimizing conditions. When a melamine resin colloidal spherical particle (SP) with tangential surface anchoring is immersed into a nematic LC, it distorts an otherwise homogeneous director field $\mathbf{n}(\mathbf{r})$ in the surrounding bulk, forming what is known as an elastic quadrupole (19,23-26,41). The quadrupolar configuration of $\mathbf{n}(\mathbf{r})$ around SPs (23) arises from the mismatch of the preset homogeneous far-field director $\mathbf{n}_0$ in the LC cell and the tangential alignment of LC molecules at the SP's surface. Arising from the minimization of free energy, the ensuing director distortions are axially symmetric with respect to an axis parallel to $\mathbf{n}_0$ and have also a mirror symmetry with respect to the plane normal to $\mathbf{n}_0$ and passing through SP's equator. There are two surface point defects called "boojums" located at the opposite poles of SP along $\mathbf{n}_0$ (23,25) and facing the substrates in a homeotropic cell (Figs. 1*A*, 2*A* and 3*A*), where $\mathbf{n}_0$ is set normal to substrates by aligning surface treatment at confining plates (*Materials and Methods*). As a consequence of a thickness of the homeotropic cell $h = 5.5$ μm in our experiments being only slightly larger than a diameter $D_0 = 2R_0 = 3$ μm of SPs, the director distortions induced by particles are partially suppressed by a strong anchoring at confining plates and do not propagate over large distances as in a bulk nematic LC (23,51). Furthermore, the interaction of such particles with confining substrates can be also understood by invoking analogies with electrostatics, where strong boundary conditions prompt the emergence of effective "image quadrupoles" (38) and lead to particles repelling from the confining substrates. Typically, a balance between elastic repulsion from the walls of the confining substrates and gravitational force $F_g = (4/3)\pi R_0^3 \Delta\rho g \approx 0.07$ pN, where $g = 9.8$ m s$^{-2}$ is gravitational acceleration and $\Delta\rho \approx 500$ kg m$^{-3}$ is a difference between densities of melamine resin SPs and LC, determines a vertical position of colloidal particles in LC cells. However, due to the tight confinement in our thin LC cells, such small symmetry-breaking contribution of $F_g$ can be neglected so that the SP's vertical position is determined mainly by a force of elastic repulsion from the confining walls, which can reach tens of piconewtons (35) with a corresponding repulsive potential from hundreds to thousands of $k_B T$ (here $T$ is temperature and $k_B$ is the Boltzmann constant) (38,51). Thus, elastic forces prevent sedimentation of SPs, so that they stay centered between two substrates as the elastic repulsion from both walls is about equal due to a symmetry of quadrupolar $\mathbf{n}(\mathbf{r})$-distortions around the particle (Figs. 1*A*,3*A*). Tangential boundary conditions at SPs were confirmed by optical microscopy observations (Fig. 1*B*-*D*), where the quadrupolar $\mathbf{n}(\mathbf{r})$-distortions can be seen between crossed polarizers as SPs encircled by a sequence of four bright lobes separated by four narrow dark regions (Fig. 1*C*). A full wave ($\lambda$ = 530 nm) phase retardation plate with a slow axis $\gamma$ placed after a sample at 45° between crossed polarizers (Figs. 1*D*, 2*B*) allows for determining the azimuthal orientation of the in-plane projection of a tilted director at the SP's surface based on the interference colors in an experimental texture. Interference colors within the LC texture are blue or yellow when the in-plane projection is respectively parallel or perpendicular to $\gamma$. The sequence of interference colors in the experimental textures (Fig. 1*D*) confirms that the projections of the tilted director are distributed around SP accordingly to planar boundary conditions. The calculated polarizing microscopy textures without and with a retardation plate (Fig. 3*D*,*E*) based on the calculated quadrupolar configuration of $\mathbf{n}(\mathbf{r})$ (Fig. 3*A*) are in a good agreement with experimental observations (Fig. 1*C*,*D*).

In the equilibrium state, at no electric fields applied, elastic quadrupoles (Figs. 1*A*, 3*A*) drift freely in the plane of the cell due to the Brownian motion (Fig. 1*G*) with displacements independent of the in-plane direction (Fig. 1*H*), which confirms that a slight rubbing of alignment layers along $\mathbf{e}_r$ (Fig. 1*A*) in our experimental cells does not impede the directionality of particles free diffusion. The vertical displacements are suppressed due to a strong elastic repulsion from both substrates. The simplest motion of a Brownian SP can be described by its mean square displacement (MSD) as a function of time MSD=$4D_{tot}t^{\alpha}$ with a diffusion coefficient $D_{tot}$ (Fig. 1*I*). Even though an isolated particle in a dilute SPs-LC dispersion shows a linear self-diffusion with $\alpha$=1 (Fig. 1*I*), its lateral motion is strongly hindered by the tight confinement with a gap of $h\approx1.8D_0$. Using optical microscopy tracking of the Brownian motion of an isolated SP, one can determine a diffusion coefficient $D_{tot}$=7.55×10$^{-4}$ μm$^2$ s$^{-1}$ of in-plane diffusion (23,42,52), which is more than 2.5 times smaller than a diffusion coefficient $D_{calc}$=19.5×10$^{-4}$ μm$^2$ s$^{-1}$ calculated for a similar but defect-free particle moving in an isotropic liquid with a viscosity $\eta$=75 mPa s of a LC using the Stokes-Einstein expression $D_{calc}=k_BT(6\pi\eta R_0)^{-1}$ (53). This significant difference shows that the strong surface anchoring at confining walls in a thin cell gap decreases the mobility of SPs because of the strong elastic coupling of

**n(r)**-distortions around SPs to confining substrates. In concentrated SPs-LC dispersions, particles tend self-arranging into two-dimensional arrays (Fig. 1$F$) with inter-particle distances determined by the number density of SPs and elastic repulsion between elastic quadrupoles (Fig. 1$J$). Their lateral displacements become confined in transient cages formed by interactions with neighboring particles, which prevents them from diffusing freely throughout the sample. Thus, the thermal motion of a single SP in a concentrated dispersion becomes sub-diffusive with $\alpha$<1 (Fig. 1$I$). A mean square displacement of SPs in concentrated dispersions is influenced not only by confinement between walls but also by elastic interactions with neighbors leading to a collective diffusion (54,55). A pair of isolated elastic quadrupoles in a dilute dispersion experience repulsive interactions which keep them apart at some distance around 12 $\mu$m or larger due to elastic repulsion (Fig. 1$E$ and 1$J$).

**Locomotion of Colloidal Spheres.** We observed that upon applying voltage with a square waveform shown in Fig. 2$D$, SPs start a bi-directional locomotion along a rubbing direction $\boldsymbol{e}_r$ (Fig. 2$B$-$E$, Movies S1 and S2). Due to a negative dielectric anisotropy $\Delta\varepsilon$=−4.8 of a used nematic LC, the director **n(r)** tends to reorient orthogonally to an electric field applied across the cell (Fig. 2$A$). Therefore, if an applied voltage $U$ that is larger than a threshold voltage $U_{th}$ of the Fréedericksz transition (29), the director far from the particle tilts away from its initial homeotropic orientation by an angle depending on the voltage value (Fig. 2$A$,3$F$). It is important to mention that preexisting deformations of **n(r)** around SP make the tilt of **n(r)** threshold-free so that further reorientation of the pre-deformed director structure is thresholdless. In our system, the direction of a **n(r)**-tilt is macroscopically homogeneous because of rubbing (56) along $\boldsymbol{e}_r$ on the top substrate $\boldsymbol{e}_{r,t}$ and on the bottom substrate $\boldsymbol{e}_{r,b}$ (Figure 2$A$), which also effectively "smoothens" the threshold-like switching one would expect far from particles for purely perpendicular boundary conditions. The rubbing-defined direction of **n(r)**-tilt along $\boldsymbol{e}_r$ was confirmed using conoscopic observations (57).

An applied voltage used in experiments has a square waveform (Fig. 2$C$,$D$) and with an amplitude alternating between $U_1$ and $U_2$ peak values at low modulation frequency $f_m$ and a 50% duty cycle, with $0<U_2/U_1\leq0.5$ (Fig. 2$D$). To make sure that observed effects are unrelated to ion electrokinetic transport, and to guarantee that we avoid effects associated with transport of ions at applied electric fields of low-frequency $f_m$, we also used a high carrier frequency $f_c$=1-10 kHz. When, during the first half-cycle of a square wave, a voltage $U_1 > U_{th}$ is applied across the LC cell (Fig. 2D), $\mathbf{n(r)}$ tilts along $\mathbf{e}_r$ away from $\mathbf{n}_0 \| z$ and towards the $y$-axis (Fig. 2A,3F). The deformed configuration of the director around the particle loses the quadrupolar symmetry, whereas a resulting nematoelastic multipole morphs. This can be observed from the polarizing microscopy images, especially with a retardation plate: note the change in the sequence of yellow and blue colors around the particles without and with voltage applied (compare Figs. 1D and 2B), which is in agreement with numerical modelling (compare Figs. 3E and 3J). Effectively, visible distortions areas around SPs become larger, as can be seen from microscopy micrographs in Fig. 2B (compare Figs. 1C,D and 2B). The evolution of director distortions around the particles is also confirmed by theoretical calculations (compare Figs. 3B,C and G,H). The nonreciprocal director evolution during this switching prompts the particle to move either along $y$ or -$y$ direction with velocity $v_1$. In the other half-cycle of a square wave, when the voltage is changed to $U_2 \leq U_1/2$, the director tends rotating back to $\mathbf{n}_0 \| z$ and nematoelastic multipole morphs/rotates accordingly back to the original orientation. This also drives particles to move along $\mathbf{e}_r$ but in the reverse direction back to the original position with velocity $v_2$. The reorientation of the director depends on the strength of the electric field (29) and in the first half-cycle, when $U_1 > U_{th}$, it is faster than in the second half-cycle at a lower effective voltage ($U_2 \leq U_1/2$). The velocity $v_2$ is smaller in the second half cycle of the applied signal when $U_2 \leq U_1/2$ as can be seen from the experiment (Fig. 2E). The angular trajectories of electrically-driven and elastic-relaxation-driven director rotations within the two parts of the voltage modulation period are different. As a result, such non-reciprocal reorientation of $\mathbf{n(r)}$ in both half-cycles causes a difference in magnitude and velocity of the forward and reverse displacements and a full cycle gives some small net forward displacement. The repeated periodic switching between $U_1$ and $U_2$ results in a directional net "squirming"-like (30-34) motion of SPs along $\mathbf{e}_r$ (Fig. 2C-E, Movies S1 and S2). The used voltage driving scheme is designed to produce an asymmetric deformation and maintain the director field asymmetry while the amplitude of the applied electric field is modulated to yield the squirming motion of the elastic colloidal multipole.

We characterized the particles' motion using tracking video microscopy. Tracking SPs and measuring velocity $v_1 \approx 3.6$ μm/s in the first half-cycle of the applied voltage (Fig. 2E), a propulsion force $F_1$ pushing a particle forward along $\mathbf{e}_r$ can be found by balancing it (because of low Renolds number) with a Stokes force as $F_1 = (k_B T / D_{tot}) v_1 \approx 20$ pN. During the second half-cycle with $U_2 < U_1/2$, $\mathbf{n(r)}$ reorients back to an initial orientation along $\mathbf{n}_0$ and a particle is pushed back towards the original position with a force $F_2 = (k_B T / D_{tot}) v_2 \approx 7$ pN. The periodic forward and reverse displacements of SP with different velocity $v_1 > v_2$, periodic change of the effective volume of nematoelastic multipoles (Fig. 3$B,C,G,H$) and non-reciprocal relaxation of $\mathbf{n(r)}$ during two half cycles of a square-wave signal result in a net directional "squirming" motion (30-34) along $\mathbf{e}_r$ with a velocity $v_{net}$, which can reach up to 1-2 µm/s (Fig. 2$C,E,F$) depending on parameters of applied voltage. Apparent or net velocities of SPs are low enough that the Ericksen number Er=$\gamma v_i R_0 / K$ is much smaller than 1, where and index $i$ indicates forward, reverse or net motion. This small Er means that $\mathbf{n(r)}$ is not modified by a flow of a LC fluid during their directional motion. Because of symmetry of $\mathbf{n(r)}$-distortions with respect to the equator of SPs (Fig. 2$A$), in dilute samples where particles are far apart from each other, they can select to move in either directions $y$ or -$y$, along $\mathbf{e}_r$ in Fig. 2$A,B$. The initial displacement either along $y$ or -$y$ direction is selected by SPs arbitrarily upon the initial switching a voltage on and it determines a further SPs' motion direction. The rubbing $\mathbf{e}_r$ does not introduce significant pretilt into the homeotropic alignment of $\mathbf{n(r)}$, which can be verified by lateral displacements independent of the in-plane direction (Fig. 1$H$) and conoscopic microscopy. Therefore, at the moment when the square-wave voltage with a fast rise-time is turned on, there is no strongly defined or preferred azimuthal direction of a tilt for $\mathbf{n(r)}$ and it is tilting in multiple directions especially in the area close to the particle as $\mathbf{n(r)}$ was already pretilted there radially due to the quadrupolar symmetry of distortions around it. The preferred homogeneous tilt along $\mathbf{e}_r$ establishes within several first cycles after the square-wave voltage being applied. The tilting dynamics at the moment of switching the voltage on can be confirmed by conoscopic observations. This symmetry-breaking event is responsible for the locomotion direction selected by a particle at the moment of turning the voltage on. Small irregularities at the particle's surface, apparent small differences of the director field near the opposite poles of the particles due to Brownian translational displacements across the cell and angular displacements at the moment of turning the voltage on can also contribute to breaking the symmetry and determining the direction of motion for each particle. The net velocity $v_{net}$ shows a maximum at the frequency $f_m$ of about two Hertz (Fig. 2$F$). Decreasing the frequency from the one corresponding to the maximum $v_{net}$ allows enough time for the tilted director to relax nearly completely to the preset vertical alignment and a particle to move back to the original position, which decreases a net displacement and $v_{net}$. Decreasing of $v_{net}$ at higher frequencies is caused by decreasing the time for relaxation of $\mathbf{n(r)}$ during either half-cycle of the applied oscillating voltage and as a result decreasing the forward and reverse displacements of particles and difference between them.

Turning the modulated voltage off completely allows for re-starting the selection of particle's locomotion direction. However, we found that the velocity directionality can be also altered during the motion via a kinetic out-of-equilibrium process even without fully turning off the modulated voltage. Figures 2*G-I* show that the locomotion direction can be changed (Movie S3) by decreasing and increasing $U_2$. First, when a waveform with $U_2/U_1=1/3$ is used, a particle moves in one direction (Fig. 2*G,H*). Decreasing $U_2$ causes the particle to slow down and stop the directional locomotion when $U_2=0$ ($U_2/U_1=0$); it just makes small, about equal displacements in both directions with no net locomotion (Fig. 2*C*). Upon further change of the waveform to the state when again $U_2/U_1=1/3$, the particle restarts directional locomotion but it may select to move in the reverse direction (Fig. 2*G,I*). The new direction depends on the particle's state and displacements direction at the moment when the applied voltages ratio becomes $U_2/U_1=1/3$ again. Thus, one can keep changing the locomotion direction of the particle by changing voltage waveforms applied to the LC cell (Fig. 2*G* and Movie S3). Importantly, as can be checked with applying waveforms with high carrier frequency like in Fig. 2*D*, the dynamic physical behavior of the nematic colloidal system is nonpolar, with motion direction generally insensitive to voltage polarity, and stems from nonreciprocal nature of LC's director evolution in applied time-varying electric field.

Details of periodic voltage induced director transformations around the particles leading to their squirming-like motion are revealed by numerical calculations (Fig. 3). The near-particle region of distortions around a particle in the tilted $\mathbf{n(r)}$ increases at the applied voltage (compare Figs. 3*B,C* and 3*G,H*). The distortions, initially axially symmetric without a voltage applied, also lose their axial symmetry because the tilt of the director at the surface of the particle with respect to the electric field induced tilt of the director in the surrounding bulk is different around the particle. In the images showing $\mathbf{n(r)}$ deviations (Fig. 3*G,H*) at applied electric field, the red contours show areas where the tilt of the director at the surface of the particle is opposite to the director tilt induced by the electric field in the surrounding bulk. This leads to increased effective volume of the distortions around the particle. The observed in experiments and numerical calculations periodic change of effective shape and size of the nematoelastic multipole around particles and non-reciprocal relaxation of $\mathbf{n(r)}$-distortions at $U_1$ and $U_2$ states result in a squirming-like locomotion of particles (Fig. 2). During the periodic change of the director tilt in the bulk the nematoelastic multipoles transform accordingly to the changing $\mathbf{n(r)}$. Even the general behavior of nematoelastic multipoles with increasing voltage is similar for particles with strong and weak anchoring (Fig. 3$K$,$L$), the strength of nematoelastic multipole moments is smaller for the particles with weak surface anchoring (Fig. 3$L$). This difference is caused by director distortions decay due to LC elasticity and action of the electric field. It is interesting to note that one can observe small changes in multipole moments even before electric field reaches a critical value of about 1.1 V. This finding is related to the fact that even without a voltage applied, there are preexisting deformations of $\mathbf{n(r)}$ around the particle, which makes it threshold-free and further reorientation is thresholdless and happens even at small voltages. The calculated corresponding polarizing microscopy micrographs (Fig. 3$D$,$E$,$I$,$J$) are in an excellent agreement with experimental observations (Figs. 1$C$,$D$ and 2$B$) and detect well the corresponding director transformations and change of the color sequence at off (compare Figs. 1$D$ and 3$E$) and on (compare Figs. 2$B$ and 3$I$,$J$) states.

**Out-of-Equilibrium Interactions of Nematoelastic Multipoles.** Pair interactions of elastic quadrupoles embedded in an infinitely large homogeneous LC depend on the orientation of a separation vector $\mathbf{r}_{cc}$ with respect to $\mathbf{n}_0$ (23). If there is no external voltage applied, elastic quadrupoles in our system stay well separated (Fig. 1$E$,$F$) even in the concentrated dispersions due to the elastic repulsion between them as $\mathbf{r}_{cc} \perp \mathbf{n}_0$ (Fig. 4$A$). When the electric field is applied and SPs are set to motion, as was described above, quadrupolar $\mathbf{n(r)}$-distortions around SPs transform into more complex nematoelastic multipoles (Fig. 3), so that they start to interact attractively and form pairs and chains of multiple particles (Fig. 4$A$, Movie S4). The external field drives our system out of equilibrium while changing repulsive interactions into dynamic attractive interactions between particles. Under the effect of an applied electric field, director tilts making nematoelastic multipoles rotate as well. However, in the tilted nematic LC in a homeotropic cell $\mathbf{r}_{cc}$ between nematoelastic multipoles is not orthogonal with the director any longer and effectively makes an angle $\theta_e<90°$ with local $\mathbf{n(r)}$, prompting them to attract when the amplitude of the applied voltage is larger than $U_{th}$. Optical microscopy textures (Fig. 2*B*) reveal that nematoelastic multipoles are rotating following the director tilt (compare the change of orientation of blue and yellow lobes in textures with a phase retardation plate in Figs. 1*D* and 2*B*). In a nematic LC, elastic pair interactions between two elastic quadrupoles are strongly repulsive when $\theta_e=0°$ and $90°$ and strongly attractive when $\theta_e=35°$-$45°$ but weak attraction is also present at about $\theta_e=20°$ or $70°$ (19,23). Therefore, it is possible to estimate that the tilt of the director during application of the voltage in the cell midplane is somewhere between $20°$ and less than $45°$ from the vertical orientation, which also can be verified via the conoscopic observations. Figure 4*C* shows decreasing of $r_{cc}$ between two nematoelastic multipoles with time under the applied voltage. After forming a pair or chains, SPs continue the directional locomotion without changing the direction along $\mathbf{e}_r$ (Fig. 4A, Movies S4 and S5) with a speed $v_{tot}$, which is an average of speeds of individual particles.

When an applied voltage is switched off, the pairs and chains of nematoelastic multipoles formed in the out-of-equilibrium regime stop moving, transform back to quadrupolar-like configurations and disassemble (Fig. 4*A*, Movies S4 and S5). This is because $\mathbf{n(r)}$ relaxes back to the vertical alignment and an angle between $r_{cc}$ and $\mathbf{n}_0$ again becomes close to $\theta_e=90°$, corresponding to the strong elastic repulsion between quadrupoles (Fig. 4*B,C*). Figures 4*C* and 4*D* show, respectively, the increasing of the separation between particles with time and decreasing of an interaction force of elastic repulsion $F_{int}$ with distance. According to the multipole expansion models (23,58), a force of elastic interactions between repelling elastic quadrupoles follows a power law behavior $F_{int} \propto r_{cc}^{-6}$, with the proportionality coefficients depending on the LC elastic constants and anchoring at the particle's surface. The experimental dependance of $F_{int}$ on separation $r_{cc}$ can be fit with a power-law exponent of $6.0\pm0.2$ (Fig. 4*D*) and corresponding separation dependence (Fig. 4*C*) can be fit well with $r_{cc}=(r_0^7+7\beta t)^{1/7}$ for the separation $r_0=3.7$ μm at time $t=0$ s and $\beta=6.05\times10^{-3}$ μm$^7$ s$^{-1}$ (27,39). The force, right after elastic quadrupoles started to repel from each other, was found to be about 6 pN, which corresponds to energy of about 1,500 $k_B T$ comparable to those measured for spherical colloids of comparable size in other works (12,19,20,25,36,37,39). The possible deviation from the power law dependence of the interparticle forces may be caused by the confinement because long-range SP-induced deformations of the director are suppressed due to the strong anchoring on the substrate walls (51).

**Collective Locomotion of Colloidal Spheres.** As was described above, because $\mathbf{n(r)}$-distortions are symmetric with respect to the equator of SPs (Figs. 2$A$, 3$F$), when elastic quadrupoles are far apart from each other in dilute samples, the direction of locomotion along $\mathbf{e}_r$ is selected by individual SPs arbitrarily upon switching voltage on. Some SPs move along $\mathbf{e}_{r,t}$ ($y$) and some along $\mathbf{e}_{r,b}$ (-$y$) directions. The situation changes drastically in dispersions of the higher number density when SPs are surrounded by multiple neighbors at relatively close distances (Figs. 1$F$). In concentrated samples, SPs start "sensing" presence of the neighbors through interactions caused by orientational elasticity of LC (Fig. 1$I$,$J$). Even though, distortions of $\mathbf{n(r)}$ around SPs are suppressed (screened) due to the surface anchoring in the tight confinement (51), SPs start interacting in our samples at relatively large distances of about 10-20 μm (Figs. 1$E$,$F$, 4$C$, 5, Movie S5). Therefore, upon switching a voltage on in concentrated samples, SPs select the same direction collectively and all move in that same direction due to elastic interactions (Fig. 5, Movie S5) and electric field driven out-of-equilibrium interactions (Fig. 4). The majority of particles move with a narrow distribution of speeds (Fig. 5$C$), depending on voltage and other parameters, with few outliers due to sticking to the surface, imperfections or aggregation. Chains of multiple SPs move with an average speed of individual particles in a chain, which sometimes can be seemingly faster than that of surrounding single particles. In dilute samples, with particles far apart, colloids can move in two opposite directions, $y$ or -$y$ (Movie S2). In the concentrated regime (Fig. 1$F$), all particles, due to synchronization via elastic interactions, move in a single spontaneously selected direction, which resembles the flocking behaviour (1-3). At any time, the amount of the order and directionality of collective locomotion of particles in our electric field driven system can be measured by an average velocity as the instantaneous velocity order parameter $v_a(t) = \left| \sum_j v_0 \exp\left(i\varphi_j(t)\right) \right| / N v_0$, where $v_0$ is a speed of a particle and $\phi_j$ is an instantaneous angle between $\mathbf{e}_r$ and the direction of motion of the $j$th particle. In our system, all SPs move in one direction with a high order of $v_a \approx 0.87$ (Fig. 5$E$,$F$, Movie S5). After the driving voltage is switched off, SPs repel and disassemble from chains moving from each other to distances determined by elastic repulsive interactions and a number density of SPs (Fig. 5*D*).

Differently from the collective schooling-like squirming motion of skyrmions in chiral LC systems (33,34), our colloidal particles can move at lower frequency $f_m$ (Fig. 2*F*) while the direction of locomotion can be reversed by restarting an applied voltage or changing the waveform of an applied voltage between $U_2/U_1 \leq 0.5$ and $U_2/U_1 = 0$ ($U_2 = 0$) (Fig. 2*G-I*) unlike by changing the frequency for skyrmion motion (59).

**Discussion and Conclusions**

In summary, we have experimentally studied electrically powered dynamic properties of colloidal particles with tangential boundary conditions dispersed in a homeotropic nematic LC cells of thickness just slightly larger than the diameter of colloids. Our findings show such particles can be set into bi-directional locomotion under the applied oscillating electric field of special waveforms. The net "squirming"-like locomotion of particles is caused by periodic transformations of the surrounding director field and resulting forward and backward displacements of particles with different velocity and non-reciprocal relaxation of the surrounding director during two half cycles of a pre-designed oscillating electric field. Moreover, driving voltage also controls the out-of-equilibrium anisotropic elastic interactions between particles, which elastically repel under equilibrium conditions.

We also show that at higher number density of particles, when they "sense" the presence of neighbor particles through interactions mediated by LC elasticity they show a collective emergent behavior in that there is a transition from bi-directional locomotion to a unidirectional locomotion. In a "collective" mode, under the applied electric field, colloids collectively select the same direction of locomotion with a high velocity order. As important feature for using in micro delivery and transportation, during the observed collective locomotion, particles in our system do not exhibit jamming or clogging (45,46) normal to the direction of locomotion preventing colloids from continuous motion and reaching the destination.

Our findings can be used for the development of different driven, active (47,48), and other out-of-equilibrium self-reconfigurable systems, microrobotics (49,50) and other technological applications. They also provide new insights into nematic LC colloids and their self-assembly, out-of-equilibrium and collective behavior. While our focus in this work was on spherical colloids, the out-of equilibrium behavior of nematic colloids can be further enriched by using particles with various geometric shapes that can be also electrically rotated around multiple axes of symmetry breaking (60), thus potentially allowing for different directionality of motions within a uniformly aligned nematic host.

**Materials and Methods**

**Sample Preparation.** A room temperature nematic LC ZLI-2806 (EMD Electronics) with negative dielectric anisotropy $\Delta\varepsilon$=-4.8 was used in our experiments as a host medium for colloidal particles. We used melamine resin spherical particles with a diameter $2R_0$=3 μm (Duke Scientific), which provide tangential surface boundary conditions for LC molecules without additional surface treatment. Colloidal dispersions with different number density of colloids were obtained by dispersing particles in a LC host at room temperature either via solvent exchange or mechanical mixing. Prepared colloidal LC dispersions were filled into LC cells while in a LC state after ~5 min sonication breaking apart pre-existing aggregates. Glass substrates with transparent indium-tin-oxide (ITO) electrodes were used for assembling LC cells. Homeotropic surface boundary conditions at confining glass substrates were set by thin films of spin-coated polyimide SE1211 (Nissan Chemical Industries, Ltd.). Polyimide-coated substrates were slightly rubbed 1-2 times with a velvet cloth applying a weak pressure of ~380 Pa along a direction $e_r$ (Fig. 1$A$) to break the azimuthal degeneracy of the director's tilt during realignment of homeotropically aligned nematic with negative dielectric anisotropy when a voltage is applied between substrates. Then substrates with rubbed alignment layers were assembled in an antiparallel fashion so that $e_r$ on top and bottom substrates are parallel but pointing in opposite directions (Fig. 1$A$). Glass spacers of diameter 5.5 μm dispersed in ultraviolet-curable glue were used to set a cell gap thickness. To minimize spherical aberrations in optical microscopy experiments involving high numerical aperture (NA) immersion oil objectives, one of two substrates was 0.15-0.17 mm thick (SPI Supplies).

**Driving voltage, Optical Microscopy and Data Acquisition.** A driving voltage was applied to ITO electrodes of a nematic cells. To generate different driving schemes and waveforms of voltage applied to LC, we used an analog function generator GW Instek GFG-8216A or a data acquisition board (NIDAQ-6363, National Instruments) and a homemade MATLAB-based software. They were used to obtain waveforms of low frequency $f_m$ voltage with a high carrier frequency $f_c$=1-5 kHz. High carrier frequency was used to avoid effects associated with a transport of ions at applied voltage of low frequency. The homemade software allowed tuning various parameters of applied voltage. Generated waveforms were controlled using an oscilloscope Tektronix TDS 2002B.

Experimental observations and data acquisition were performed using optical bright-field and polarizing microscopy imaging modalities of an inverted Olympus IX81 microscope with 20× (NA=0.4) and oil 100× (NA=1.4) objectives. Dynamics of LC textures and translational motion of colloidal particles was recorded using a charge-coupled device (CCD) camera (Flea, PointGrey) at a rate of 15 or 30 frames per second and the exact spatial positions of colloidal spheres as a function of time were then determined from captured video using motion tracking plugins of the ImageJ (freeware from NIH) analyzing software.

**Numerical Modelling**. Nematic director configurations and distortions around spherical particles (Fig. 3) were computationally modelled based on Landau-de Gennes free energy minimization using a home-built Matlab program. The total LC bulk free energy density with the LC continuum represented by a tensorial field read (61):

$$f_{\text{bulk}} = \frac{A}{2} Q_{ij}Q_{ji} + \frac{B}{3} Q_{ij}Q_{jk}Q_{ki} + \frac{C}{4}\left(Q_{ij}Q_{ji}\right)^2 + \frac{L_1}{2}\left(\frac{\partial Q_{ij}}{\partial r_k}\right)^2 + \frac{L_2}{2}\frac{\partial Q_{ij}}{\partial r_j}\frac{\partial Q_{ik}}{\partial r_k} + \frac{L_6}{2} Q_{ij}\frac{\partial Q_{kl}}{\partial r_i}\frac{\partial Q_{kl}}{\partial r_j} - \frac{1}{2}\varepsilon_0\Delta\varepsilon(n_iE_i)^2 \qquad (1)$$

with summation of indices implied $i,j,k$=$x,y,z$. The 3-by-3 tensorial order parameter **Q** is defined as **Q**=$S$/2(**n**⊗**n-I**) by the scalar order parameter $S$, LC director **n(r)**, and identity matrix **I** (61). The Landau-de Gennes coefficients $A$=-1.72×10$^5$ J·m$^{-3}$, $B$=-12.2×10$^6$ J·m$^{-3}$, and $C$=1.73×10$^6$ J·m$^{-3}$ were adopted (61), and the elastic constants $L_1$=6.31 pN, $L_2$=10.96 pN, and $L_6$=0.49 pN were calculated using reported Frank-Oseen elasticities of ZLI-2806 (62,63). Finally, the electric contribution was computed using the dielectric anisotropy $\Delta\varepsilon$=-4.8 of the nematic LC, vacuum permittivity $\varepsilon_0$=8.85×10$^{-12}$ F·m$^{-1}$, and an electric field **E** assumed to be uniformly vertical across the numerical volume. At the boundary of nematic LC, including particle-LC interfaces and cell substrates, a surface energy density took the form (64):

$$f_{\text{surf}} = W_0 \left( P_{ik}\widetilde{Q}_{kl}P_{lj} - \frac{3}{2}S\cos^2(\varphi)\, P_{ij} \right)^2 \qquad (2)$$

with $\mathbf{P}=\mathbf{v}\otimes\mathbf{v}$ the surface projection tensor, $\mathbf{v}$ the surface normal vector, and $\widetilde{Q}=S/2(\mathbf{n}\otimes\mathbf{n})$. The surface anchoring coefficients $W_0 = 10^{-5}$ J m$^{-2}$ and $10^{-3}$ J m$^{-2}$ for a particle surface were used in calculations with surface anchoring angle $\varphi=90°$ (for spheres with tangential anchoring) and $0°$ (for substrates with homeotropic anchoring).

For each applied voltage $U=\int E_z dz$, the LC director field $\mathbf{n}(\mathbf{r})$ was first computed in a one-dimensional box along the $z$-direction with a homeotropic anchoring at top and bottom without a colloidal particle present. The obtained $\mathbf{n}(\mathbf{z})$ subsequently served as the initial background director profile of the complete simulation in a box sized 6 $\mu$m in each dimension and with a spherical particle with diameter $D_0=3$ $\mu$m located at the center of numerical volume. Periodic boundary conditions were implemented in $x$- and $y$- directions of the box. The energy-minimizing director field $\mathbf{n}(\mathbf{r})$, along with surface-induced distortions (defined as the difference between $\mathbf{n}(\mathbf{r})$ and background $\mathbf{n}(\mathbf{z})$), were then visualized using Paraview (Fig. 3). The calculation of corresponding multipole moments is detailed in Refs. 26 and 64. Here we presented the sum of absolute values $\sum_m |q_{lm}|$ for each multipole moment. Corresponding microscopy micrographs were calculated using the extended Jones matrix method.

**Author Contributions:** I.I.S. designed research and directed the project; B.S. performed experiments; J.-S. Wu performed modelling and calculations; B.S. and I.I.S. analyzed data and wrote the paper.

**Acknowledgments**

We acknowledge discussions with R. E. Adufu, M. Tasinkevych and T. Lee. I.I.S. acknowledges hospitality of the International Institute for Sustainability with Knotted Chiral Meta Matter (SKCM$^2$) at Hiroshima University, where he was partly working on this article. This research was supported by the U.S. Department of Energy, Office of Basic Energy Sciences, Division of Materials Sciences and Engineering, under contract DE-SC0019293 with the University of Colorado at Boulder.

**Figures**

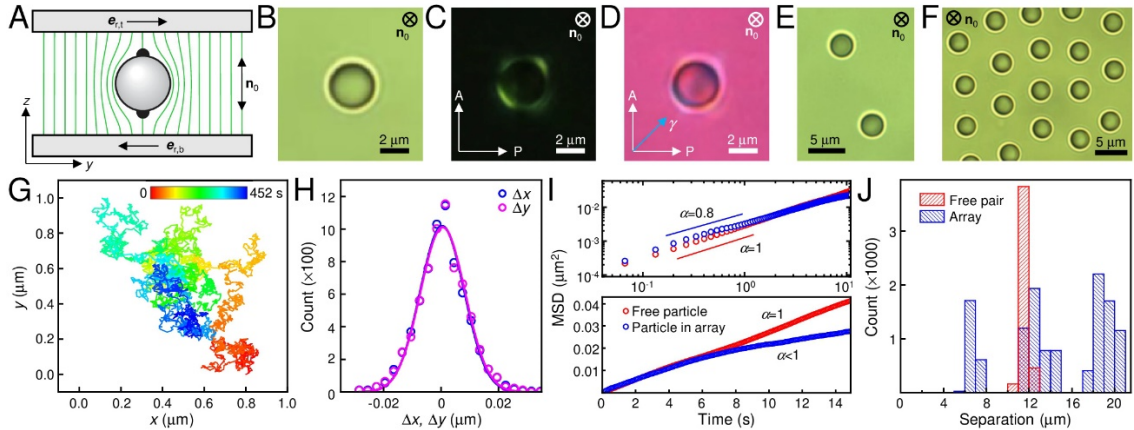

**Figure 1.** Elastic quadrupoles in a homeotropic nematic cell in an equilibrium state. (*A*) Schematic diagram of an elastic quadrupole and surrounding director field (green lines) in a homeotropic cell. Black hemispheres at the poles of a particle show a surface point defects "boojums" Top and bottom confining substrates are slightly rubbed respectively in $e_{r,t}$ and $e_{r,b}$ directions. (*B-D*) Microscopy micrographs of an elastic quadrupole obtained using (*B*) bright-field and polarizing microscopy without (*C*) and with (*D*) a phase retardation plate after the sample; A, P and $\gamma$ mark respectively crossed analyzer and polarizer and a "slow" axis of a phase retardation plate. (*E* and *F*) Bright-field microscopy micrographs showing a pair (*E*) and a 2D array of multiple (*F*) particles. (*G*) Color coded trajectory of Brownian motion of an elastic quadrupole in the plane of a homeotropic cell. Color coded bar shows an elapsed time. (*H*) Histograms of displacements in the plane of a cell along the x and y directions. Solid lines show a Gaussian fit to the data (open symbols). (*I*) Log–log (*Top*) and linear (*Bottom*) plots of mean-square displacement (MSD) measured for a single elastic quadrupole moving (*B*) freely well separated from neighbors and (*F*) within an array of elastic quadrupoles in a homeotropic cells with $h$=5.5 μm. Experimental data were fitted using the expression MSD = $4D_{tot}t^{\alpha}$ (51). Diffusion of a single free elastic quadrupole (*B*) corresponds to the Brownian motion with α = 1 but the elastic quadrupole within an array (*F*) shows subdiffusive motion (α < 1) due to suppressed mobility caused by elastic repulsion from neighbors. (*J*) Histogram of separations between elastic quadrupoles in a pair (*E*) and in an array (*F*).

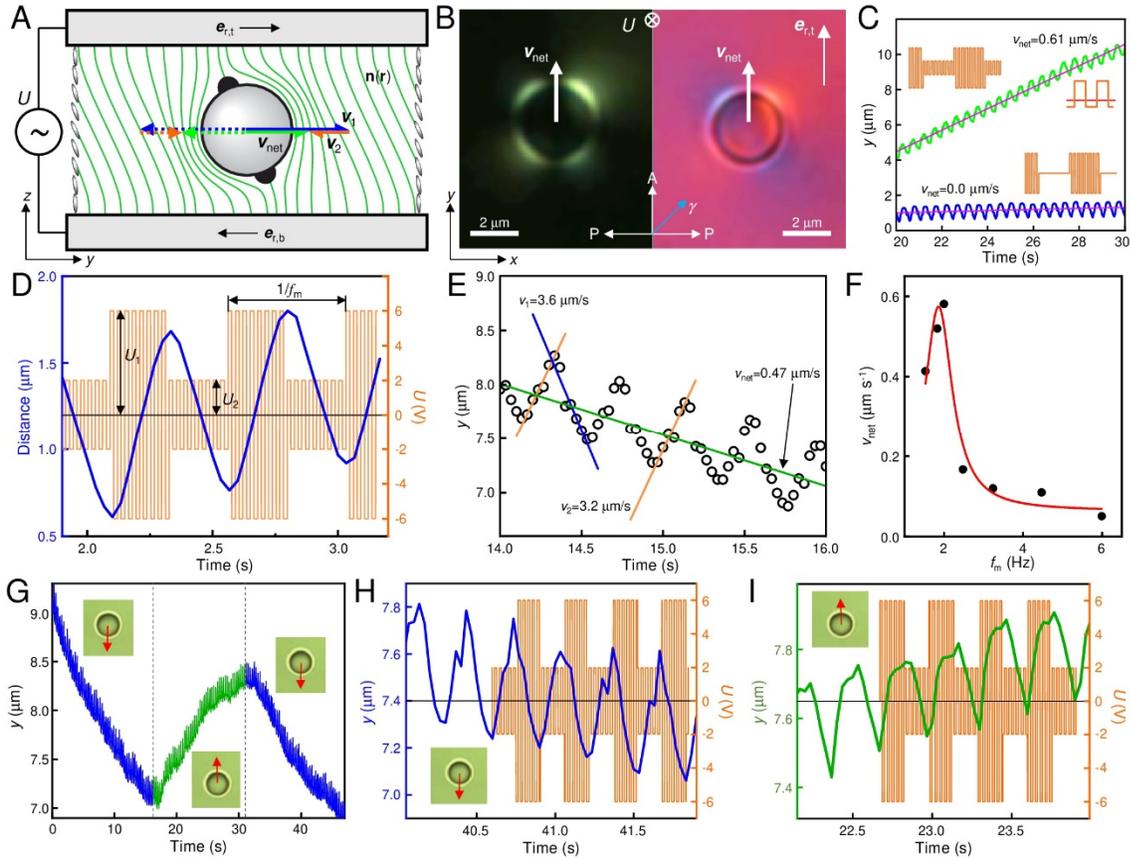

**Figure 2.** Electric field induced locomotion of colloidal particles in a homeotropic nematic cell. (*A*) Schematic diagram of a director field around a particle shown in (*B*) under an applied voltage. Ellipses represent the LC molecules. The pointing in two opposite directions solid and dashed arrows corresponding to forward $v_1$ (blue), reverse $v_2$ (green) and net $v_{net}$ (orange) displacements show that the locomotion along $\mathbf{e}_r$ is equally probable in both directions but the direction is preserved during motion after selected by a particle upon switching a voltage on. (*B*) Polarizing optical micrographs of LC textures around a particle moving under the applied voltage taken without (*Left*) and with (*Right*) a phase retardation plate. (*C*) Displacement of a particle over time at $0 \leq U_2/U_1 \leq 0.5$ and $U_2/U_1 = 0$ waveforms of the applied voltage (shown as the insets) at $f_m = 2.1$ Hz. Applying the former waveform results in the net directional locomotion of the colloidal particle and there is no locomotion when applying the latter. A waveform of applied low frequency voltage asymmetric with respect to a zero level marked by a red line in the top right inset can also result in the net directional locomotion of the colloidal particle. Solid lines are the best linear fits to experimental displacements. (*D*) Waveform of an applied voltage that induces locomotion of particles with low modulation frequency $f_m = 2.1$ Hz and high carrier frequency $f_c = 1-10$ kHz. For the purpose of demonstration, $f_c$ in a plot is represented by a lower frequency. (*E*) Plot showing the detailed displacements of a particle under the applied voltage similar as shown in (*D*) resulting in a locomotion with a net velocity $v_{net}$. Solid lines show the linear fit of trajectory fragments corresponding to forward and reverse displacements and resulting locomotion. (*F*) Dependence of net velocity on $f_m$. A red solid line is a guide to eyes. (*G-I*) Plots showing the change of locomotion direction (*G*) upon restarting the applied voltage ($f_m$=3.2 Hz). Dashed lines separate regions with opposite directions of locomotion.

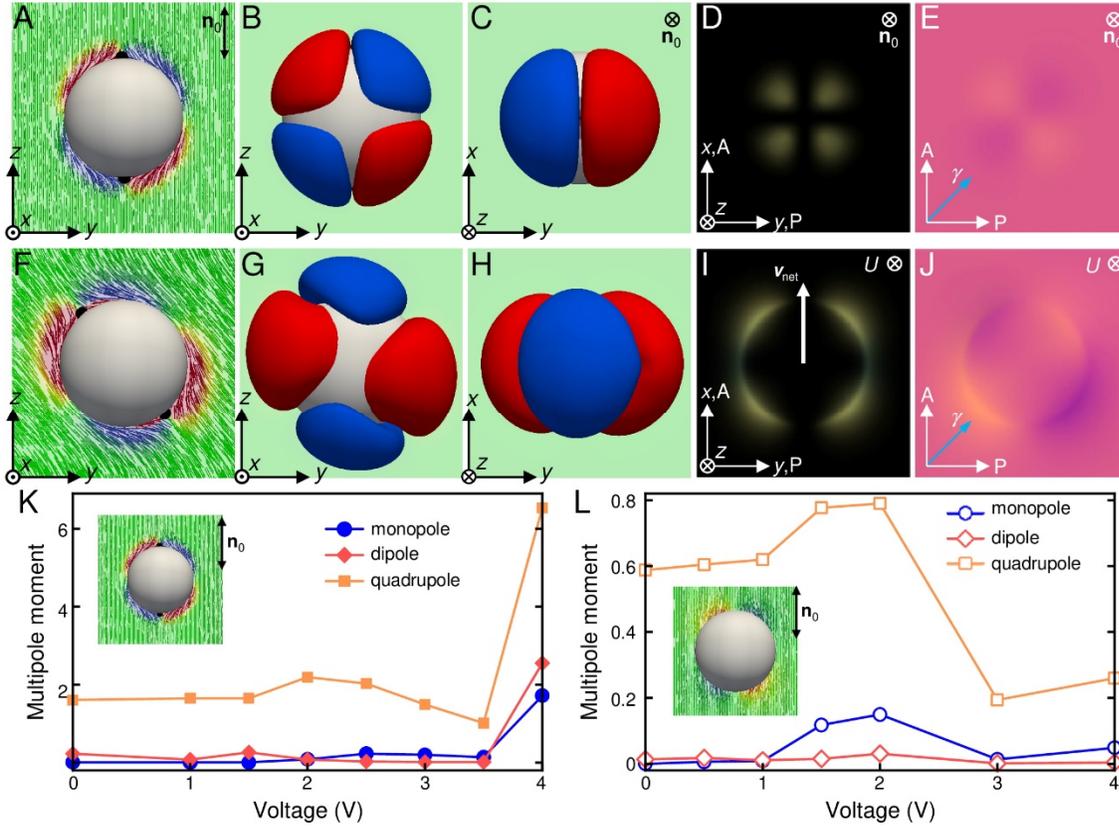

**Figure 3.** Director transformations around a particle. (*A-C,F-H*) Computer-simulated **n**(**r**) in the vertical plane *z-y* going through the middle of the particle (*A,F*) and contours marking area around the particle where the deviations of $n_y$=±0.03 (*B,C,G,F*) respectively without (*A-C*) and with (*F-H*) voltage applied. Rods in *A* and *F* represent the director field. (*D,E,I,J*) Corresponding computer-simulated microscopy micrographs for polarizing microscopy without (*D,I*) and with (*E,J*) a phase retardation plate after the sample; A, P and γ mark respectively crossed analyzer and polarizer and a "slow" axis of a phase retardation plate. (*K,L*) Calculated elastic multipole moments dependent on the applied voltage for a particle with (*K*) a strong ($W_0$=10^{-3} J m^{-2}) and (*L*) weak ($W_0$=10^{-5} J m^{-2}) surface anchoring.

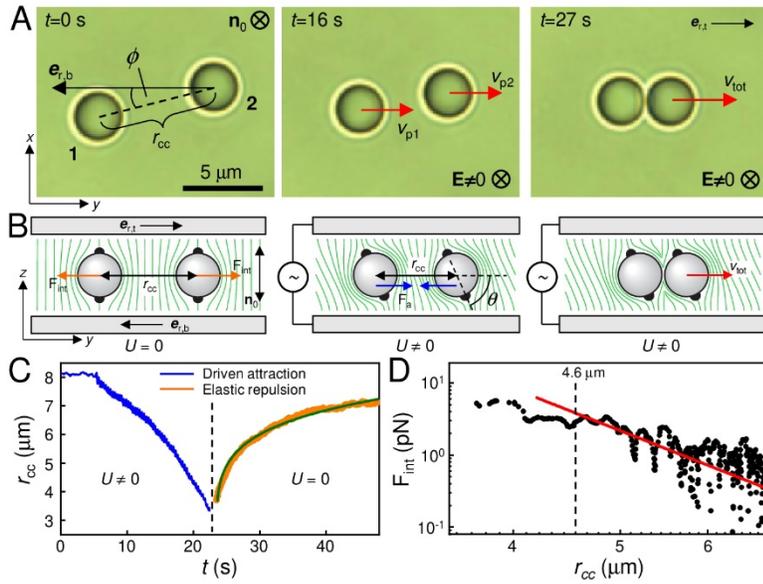

**Figure 4.** Out-of-equilibrium pair interactions of particles. (*A*) Sequence of bright-field microscopy images showing a time evolution of out-of-equilibrium attractive pair interaction between two particles in a homeotropic nematic cell upon applying the ac electric field. (*B*) Schematic diagrams showing the director field around two interacting particles corresponding to experimental images in (*A*). (*C*) Plots showing a time evolution of separation $r_{cc}$ for the pair interactions between two particles under applied voltage ($U\neq0$ at $f_m$=3 Hz) and no voltage applied. A green solid line is the best fit to experimental data of elastic repulsion between two elastic quadrupoles when an applied voltage is switched off with $r_{cc}=(r_0^7+7\beta t)^{1/7}$ for $r_0$=3.7 μm and $\beta$=6.05 10⁻³ μm⁷ s⁻¹ (27,39). A vertical dashed line separates regions when voltage is on and off. (*D*) Log-log plot of an interaction force $F_{int}$ versus separation between elastic quadrupoles corresponding to repulsive elastic interactions in (*C*). A red line is a best fit to a power-law exponent of 6.0 ± 0.2 corresponding to quadrupolar elastic interactions (19,23). A vertical dashed line indicates the boundary of the fitting region with a value indicated at the top.

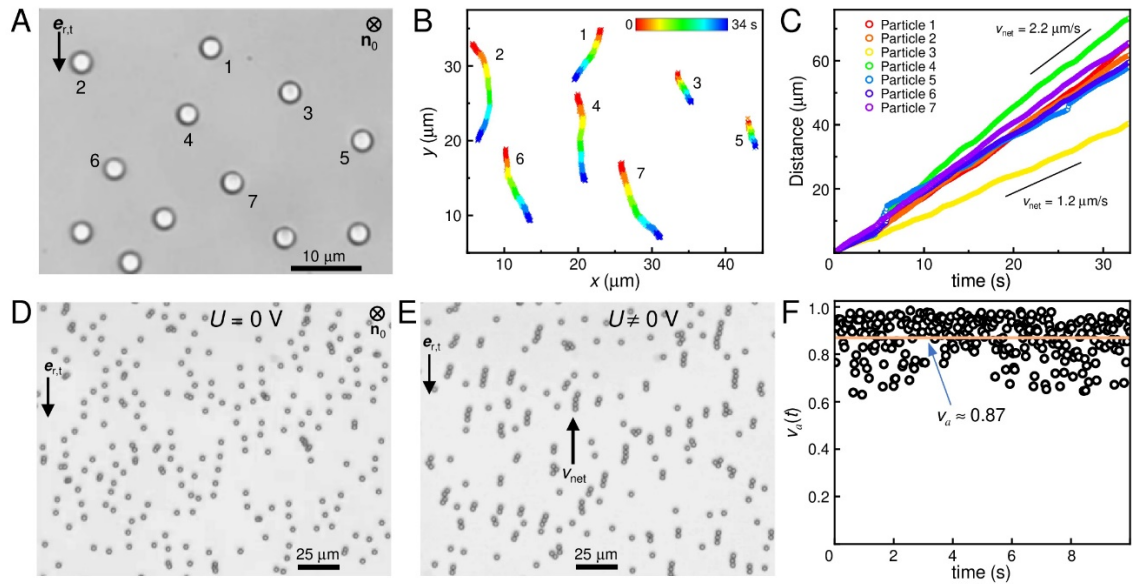

**Figure 5.** Emergent collective locomotion of particles in a homeotropic nematic cell. (*A*) Bright-field micrograph of multiple particles assembling into a 2D structure at higher number density. (*B*) Colour coded trajectories of particles shown in (*A*) moving along a rubbing direction $e_r$ when a voltage $U_1$ =6 V ($U_2/U_1$=1/3) at $f_m$=2.5 Hz was applied. Colour coded bar shows elapsed time after application of voltage. (*C*) Distance versus time showing a speed distribution of moving particles in (*A* and *B*). (*D*) Distribution of repelling particles before an application of an electric field. (*E*) Directional locomotion of particles driven by an applied voltage at the high number density. Single particles and their chains move in one direction along $e_r$. (*F*) Velocity order parameter of particles moving in (*E*) over time. A solid line is a linear fit to data (open symbols) indicating an average velocity order parameter.

# Supporting Information

**Movie S1 (separate file).** Electric field driven directional motion of two colloidal particles with tangential boundary conditions moving in the same direction in a homeotropic nematic cell at $U_1 = 6$ V and $f_m = 3$ Hz.

**Movie S2 (separate file).** Electric field driven directional locomotion of colloidal particles with tangential boundary conditions moving in the opposite directions in a homeotropic nematic cell at $U_1 = 6$ V and $f_m = 3.75$ Hz.

**Movie S3 (separate file).** Colloidal particle changing the motion direction upon the change of waveforms of an applied voltage at $U_1 = 6$ V and $f_m = 3.2$ Hz.

**Movie S4 (separate file).** Out-of-equilibrium elastic attractive pair interaction and motion of two colloidal particles under an applied electric field (at $U_1 = 6$ V and $f_m = 3$ Hz) and elastic repulsive interaction when an electric field is switched off.

**Movie S5 (separate file).** Out-of-equilibrium elastic attractive interaction and collective motion of multiple colloidal particles under an applied electric field (at $U_1 = 6$ V and $f_m = 2.5$ Hz) and elastic repulsive interaction when an electric field is switched off.